\documentclass{article}
\usepackage{graphicx}  
\usepackage{amsmath}   
\usepackage[compress]{cite}
\usepackage{amssymb}   
\usepackage{bm} 
\usepackage{dcolumn}
\usepackage{color}
\usepackage{mathrsfs}
\usepackage{amsfonts}
\usepackage{varioref}
\RequirePackage[colorlinks,citecolor=blue,urlcolor=magenta,linkcolor=blue]{hyperref}
\def\LL{Lanczos-Lovelock }

\def\gr{general relativity}

\labelformat{section}{Section #1} 
\labelformat{subsection}{Section #1} 
\labelformat{subsubsection}{Section #1}
\labelformat{subsubsubsection}{Section #1}
\labelformat{equation}{Eq.~(#1)} 
\labelformat{figure}{Fig.~#1} 
\labelformat{subfigure}{Fig.~\thefigure#1} 
\labelformat{table}{Tab.~#1} 
\labelformat{appendix}{Appendix #1}
\title{Field equations for Lovelock gravity: An alternative route}
\author{Sumanta Chakraborty
\footnote{sumantac.physics@gmail.com}\\
{\small{Department of Theoretical Physics, Indian Association for the Cultivation of Science, Kolkata 700032, India}}}
\begin{document}
  
\maketitle
\begin{abstract}
We present an alternative derivation of the gravitational field equations for Lovelock gravity starting from the Newton's law, which is closer in spirit to the thermodynamic description of gravity. As a warm up exercise, we have explicitly demonstrated that projecting the Riemann curvature tensor appropriately and taking a cue from Poisson's equation, the Einstein's equations immediately follow. The above derivation naturally generalizes to Lovelock gravity theories where an appropriate curvature tensor satisfying the symmetries as well as the Bianchi derivative properties of the Riemann tensor has to be used. Interestingly, in the above derivation, the thermodynamic route to gravitational field equations, suited for null hypersurfaces, emerge quiet naturally.
\end{abstract}
\section{Introduction}

Equivalence principle acts as the guiding lighthouse to understand how matter fields behave in a curved spacetime background. Unfortunately, there exists no such principle which helps to answer the opposite, viz. how matter fields curve the spacetime \cite{gravitation,LL2,Eddington:1924,MTW,Poisson,Wald,Hawking:1973uf}. Lack of such principle resulted in the vast landscape of various alternative gravity theories among which \gr\ remains the most useful one. Despite this large variety of alternative gravitational theories one can do better by imposing some physical requirements on the systems of interest. In particular, the restriction to the class of theories having utmost second order derivatives of the metric is a judicious one, since this helps to overcome the well known Ostrogradski instability by evading the existence of any ghost modes in the theory \cite{Chen:2012au,Motohashi:2016ftl,Klein:2016aiq}. Surprisingly enough the above criteria turns out to be a very interesting one, as it singles out a 
very specific class of unique gravitational theories, known in the literature as the \LL models of gravity \cite{Lanczos:1938sf,Lanczos:1932zz,Lovelock:1971yv,Padmanabhan:2011ex,
Padmanabhan:2013xyr,Chakraborty:2014joa}. In particular, all these models satisfy the criteria that Bianchi derivative of the field tensor identically vanishes \cite{Dadhich:2008df,Camanho:2015hea}. One can add to this list a large number of interesting properties, a few among them are --- (a) The action functional for \LL gravity can be broken into a bulk part and a surface contribution, which is closely related to the Wald entropy associated with the black hole horizon \cite{Padmanabhan:2004fq,Padmanabhan:2007en,Kothawala:2008in}; (b) The field equations in \LL gravity can be expressed as a thermodynamic identity on an arbitrary null surface \cite{Paranjape:2006ca,Akbar:2006er,Cai:2005ra,Ge:2007yu,Gong:2007md,Wu:2007se,
Kothawala:2009kc,Chakraborty:2010wn,Chakraborty:2015aja,Chakraborty:2015wma,
Hansen:2016gud,Dey:2016zka,Chakraborty:2016dwb}; (c) The difference between a suitably defined surface and bulk degrees of freedom in the context of \LL gravity can be interpreted as the evolution of the spacetime \cite{Padmanabhan:2009kr,Padmanabhan:2013nxa,Chakraborty:2014rga}; (d) The surface and bulk terms in the \LL gravity are related by a holographic relation between them \cite{Mukhopadhyay:2006vu,Dieks:2015jja}. 

In the standard textbook treatment, one first introduces the gravitational Lagrangian $L$, which is $\sqrt{-g}R$ for \gr, while the Lagrangian is a polynomial in the Riemann tensor for general \LL gravity. Variation of the action functional due to arbitrary variation of the metric, with appropriate boundary conditions \cite{Parattu:2015gga,Parattu:2016trq,Chakraborty:2016yna,Chakraborty:2017zep} leads to the corresponding field equations for gravity, namely the rule by which matter tells the spacetime how to curve (see also \cite{Deser:2009fq}). However, as emphasized earlier there is no physical principle which guides us to the precise mathematical structure of the gravitational action, what is more the field equations so obtained equates matter which is intrinsically \emph{quantum} with spacetime geometry describing gravity \emph{classically}. Despite this conceptual discomfort, there exist two additional representations of the gravitational dynamics, which are intrinsically observer dependent. The first one uses time like observers with four velocity $u^{a}$ and equates two quantities that this observer measures in the geometric as well as in the matter sector, leading to a scalar equation,
\begin{align}\label{Eq_01}
G_{ab}u^{a}u^{b}= \kappa T_{ab}u^{a}u^{b}
\end{align}
Here $\kappa$ stands for the gravitational constant in four dimensions and enforcing this equation for all observers with four velocity $u_{i}$ leads to the Einstein's equations $G_{ab}=\kappa T_{ab}$. The other approach uses null vectors $\ell _{a}$ (i.e., $\ell ^{2}=0$) and leads to the following scalar equation,
\begin{align}\label{Eq_02}
R_{ab}\ell ^{a}\ell ^{b}=\kappa T_{ab}\ell ^{a}\ell ^{b}
\end{align}
In this situation as well validity of the above expression for all null vectors $\ell _{a}$, along with two times contracted Bianchi identity and covariant conservation of matter energy-momentum tensor amounts to furnish the ten components of the Einstein's equations. Following \cite{Padmanabhan:2016eld}, in this article, we would like to derive \ref{Eq_01} and \ref{Eq_02} from a geometrical point of view for \LL theories of gravity. We would also argue why the route taken in this work is a natural one compared to the standard derivation of Einstein's equations from the gravitational action. We have organized the paper as follows: In \ref{Einstein} we briefly review the derivation of Einstein's equations, which will be generalized in \ref{Lovelock} to \LL theories of gravity. Finally we provide the concluding remarks before presenting an derivation regarding the curvature tensor associated with the \LL theory of gravity in the appendix.
\section{Einstein's equations following Newton's path}\label{Einstein}

To set the stage for \LL theories of gravity we would like to briefly review the corresponding situation in \gr, which will mainly follow from \cite{Padmanabhan:2016eld}. In Newtonian theory one describes gravity using the gravitational potential $\phi(\mathbf{x},t)$. Given the matter density $\rho(\mathbf{x},t)$, the dynamics of the gravitational field is being determined through the Poisson's equation $\nabla ^{2}\phi=(\kappa /2)\rho$. When one invokes principle of equivalence, the potential $\phi$ is readily identified with components of the metric, in particular, $g_{00}=-(1+2\phi)$. One then interprets the Poisson's equation in metric language as $-\nabla ^{2}g_{00}=\kappa T_{00}$, where matter energy density is interpreted as the time-time component of the divergence free, symmetric, matter energy momentum tensor $T_{ab}$. In order to have a relativistic generalization one naively makes the following replacement $\nabla ^{2}\rightarrow \square$. This immediately suggests to look for second derivatives of the metric which is a second rank tensor as well as divergence free, paving the way to the Einstein tensor. On the other hand, for the right hand side the natural choice being the matter energy momentum tensor $T_{ab}$, one ends up equating the Einstein's tensor with the matter energy momentum tensor resulting in the Einstein's equations (for a more detailed discussion, see \cite{Padmanabhan:2016eld}).  

However, we could have just followed Newton's path and look for relativistic generalization of Poisson's equation itself, possibly leading to a scalar equation describing gravity in general relativity. As a first step one must realize that the right hand side of Poisson's equation is intrinsically observer dependent, as it is the energy density of some matter field measured by an observer with some four velocity $u_{i}$. Given the symmetric and conserved matter energy momentum tensor $T_{ab}$, the matter energy density $\rho$ as measured by an observer with four velocity $u_{a}$ is $T_{ab}u^{a}u^{b}$. Hence the same observer dependence must continue to exist in \gr\ and the right hand side of the desired general relativistic equation for gravity should be $T_{ab}u^{a}u^{b}$ \cite{Padmanabhan:2016eld}. 

To derive the left hand side, i.e., analog of $\nabla ^{2}\phi$ we note that this requires to derive a scalar object which involves two spatial derivatives of the metric (since this is what is responsible for the $\nabla ^{2}\phi$ term in non-relativistic limit) and necessarily depends on the four-velocity $u^{i}$. The only tensor depending on two derivative of the metric that can be constructed corresponds to the curvature tensor $R_{abcd}$. In order to obtain double spatial derivatives acting on the metric one has to project all the components of the curvature tensor on the plane orthogonal to $u_{a}$ using $h^{a}_{b}=\delta ^{a}_{b}+u^{a}u_{b}$ and hence construct a scalar thereof. To keep generality, instead of looking for a timelike vector $u_{i}$, we will concentrate on a particular vector $\ell ^{a}$, with norm $\ell _{a}\ell ^{a}\equiv \ell ^{2}$. Then one can immediately introduce the following tensor,
\begin{equation}\label{Eq_03}
\mathbb{P}^{a}_{b}=\delta ^{a}_{b}-\frac{1}{\ell ^{2}}\ell ^{a}\ell _{b};\qquad \mathbb{P}^{a}_{b}\mathbb{P}^{b}_{c}=\left(\delta ^{a}_{b}-\frac{1}{\ell ^{2}}\ell ^{a}\ell _{b}\right)\left(\delta ^{b}_{c}-\frac{1}{\ell ^{2}}\ell ^{b}\ell _{c}\right)=\delta ^{a}_{c}-\frac{1}{\ell ^{2}}\ell ^{a}\ell _{c}=\mathbb{P}^{a}_{c}
\end{equation}
for which the final property ensures that it is a projector. Then one can define a projected Riemann curvature with respect to the vector $\ell ^{a}$ by projecting all the indices of the Riemann curvature tensor, leading to\footnote{By the very definition, the Riemann tensor $R_{abcd}$ has a generic form $\partial _{b}\partial _{c}g_{ad}$ in local inertial frame. Thus the above projection ensures that $g_{ab}$ has only double spatial derivatives when $\ell _{a}$ is a timelike vector. All time derivatives appearing in $\mathsf{R}_{abcd}$ are single in nature and hence vanishes in the local inertial frame all together. This being the prime motivation of introduction of this projected Riemann tensor. The same can be ascertained from the Gauss-Codazzi equation as well, see \cite{gravitation,Poisson}.}
\begin{equation}
\mathsf{R}_{mnrs}=\mathbb{P}^{a}_{m}\mathbb{P}^{b}_{n}\mathbb{P}^{c}_{r}\mathbb{P}^{d}_{s}R_{abcd}
\end{equation}
The scalar constructed out of this projected Riemann tensor becomes,
\begin{equation}
\mathsf{R}=\mathbb{P}_{mr}\mathbb{P}_{ns}R^{mnrs}=R-\frac{2}{\ell ^{2}}R_{ab}\ell ^{a}\ell ^{b}
\end{equation}
Note that we have thrown away another term involving $R_{abcd}\ell ^{a}\ell ^{b}\ell ^{c}\ell ^{d}$, since due to antisymmetry properties of the Riemann tensor the above quantity identically vanishes. Furthermore, the above scalar \emph{by} construction has only two derivatives of the metric and is completely spatial in the instantaneous rest frame of the time-like observer. Hence $\mathsf{R}$ is the relativistic generalization of $\nabla ^{2}\phi$ and thus must be equal to the corresponding matter energy density, which in this case corresponds to, $-(2\kappa/\ell ^{2})T_{ab}\ell ^{a}\ell ^{b}$, where $T_{ab}$ is the matter energy momentum tensor. Thus finally one obtains the following equation,
\begin{equation}
R_{ab}\ell ^{a}\ell ^{b}-\frac{1}{2}R\ell ^{2}=\kappa T_{ab}\ell ^{a}\ell ^{b}
\end{equation}
Surprisingly, through the above exercise we have achieved two results in one go. Firstly, for $\ell ^{2}=-1$, i.e., for unit normalized timelike vectors we get back the equation $G_{ab}u^{a}u^{b}=\kappa T_{ab}u^{a}u^{b}$. Then requiring these equations to hold for all timelike observers will lead to $G_{ab}=\kappa T_{ab}$. At this stage it will be worthwhile to mention that a similar approach as the above one, was taken in \cite{Straumann:2013} to derive the Einstein's equations. However the key difference between the approach in \cite{Straumann:2013} and in this work being use of the projection tensor $\mathbb{P}^{a}_{b}$ judiciously. For our approach (also see \cite{Padmanabhan:2016eld}) the use of projection tensor is of utmost importance in contrast to \cite{Straumann:2013}.

Finally the limit $\ell ^{2}\rightarrow 0$ would lead to the null version of the above equation, which reads, $R_{ab}\ell ^{a}\ell ^{b}=\kappa T_{ab}\ell ^{a}\ell ^{b}$. In this case besides demanding the validity of the above equation for all null vectors, it is important to use contracted Bianchi identity as well as covariant conservation of energy-momentum tensor, leading to: $G_{ab}+\Lambda g_{ab}=\kappa T_{ab}$. Note that in the second situation a cosmological constant has automatically come into existence.
\section{Newton leads way to Lovelock}\label{Lovelock}

In view of the above result, one immediately asks for the corresponding situation in Lovelock gravity: Can field equations of Lovelock gravity be derived following the above procedure? This is what we will explore in this section and shall show that one can indeed derive the field equations for Lovelock gravity following an equivalent procedure. Before jumping into the details let us mention some structural aspects of Lovelock gravity. Briefly speaking, Lovelock gravity corresponds to a class of gravitational Lagrangians which are polynomial in the Riemann curvature tensor yielding field equations which are second order in the metric. The analytical form of such a polynomial (also called a pure Lovelock term) of order $m$ involves $m$ Riemann curvature tensors contracted appropriately, such that
\begin{align}
L^{(m)}=\frac{1}{2^{m}}\delta ^{a_{1}b_{1}\cdots a_{m}b_{m}}_{c_{1}d_{1}\cdots c_{m}d_{m}}R^{c_{1}d_{1}}_{a_{1}b_{1}}\cdots R^{c_{m}d_{m}}_{a_{m}b_{m}}\equiv \frac{1}{m}P^{ab~(m)}_{cd}R^{cd}_{ab}
\end{align}
The above relation defines the tensor $P^{ab~(m)}_{cd}$ associated with the $m$th order \LL gravity, having all the symmetries of the Riemann tensor with the following algebraic structure
\begin{align}\label{Entropy_Tensor}
P^{ab~(m)}_{cd}=\frac{m}{2^{m}}\delta ^{aba_{2}b_{2}\cdots a_{m}b_{m}}_{cdc_{2}d_{2}\cdots c_{m}d_{m}}R^{c_{2}d_{2}}_{a_{2}b_{2}}\cdots R^{c_{m}d_{m}}_{a_{m}b_{m}}
\end{align}
The tensor $P^{ab~(m)}_{cd}$ satisfies an additional criteria $\nabla _{a}P^{ab~(m)}_{cd}=0$, which ensures that the field equations derived from this Lagrangian are of second order. In what follows we will exclusively concentrate on the $m$th order \LL gravity and hence shall remove the symbol ``$(m)$'' from the superscripts of various geometrical expressions.

One can now start from the Poisson's equation and fix the right hand side to be $-T_{ab}\ell ^{a}\ell ^{b}/\ell ^{2}$, which is the matter energy density associated with the vector field $\ell ^{a}$. In order to get the left hand side we must construct an appropriate curvature tensor suited for the Lovelock gravity and project it using the projector $\mathbb{P}^{a}_{b}$ introduced in \ref{Eq_03} before constructing a scalar out of it. There are two possible choices for such a curvature tensor among which we will discuss the simpler one in the following, while the complicated one is deferred to \ref{App_01}. The curvature tensor described here was first introduced in \cite{Dadhich:2008df,Camanho:2015hea} and is defined as follows,
\begin{align}
\mathtt{R}^{ab}_{cd}=\frac{1}{2}\left(P^{ab}_{mn}R^{mn}_{cd}+P^{mn}_{cd}R_{mn}^{ab}\right)-\frac{(m-1)}{(d-1)(d-2)}\left(\delta ^{a}_{c}\delta ^{b}_{d}-\delta ^{a}_{d}\delta ^{b}_{c}\right)L
\end{align}
where, $L$ is the $m$th order Lovelock polynomial and $P^{ab}_{cd}=\partial L/\partial R^{cd}_{ab}$. 
Since the $m$th order \LL Lagrangian depends on $m$ powers of Riemann, the above definition for $P^{ab}_{cd}$ ensures that it depends on $(m-1)$ powers of Riemann tensor and hence exactly coincides with \ref{Entropy_Tensor}. Furthermore, the tensor $P^{ab}_{cd}$ can be easily generalized to the full Lovelock polynomial by just adding over different $m$ values, but we will concentrate on a single term in the full \LL Lagrangian. Also note that the above defined `Lovelock' Riemann tensor has all the symmetries of the original `Einstein' Riemann tensor $R^{ab}_{cd}$. Further it satisfies the contracted Bianchi identity, which will be sufficient for our purpose. Given the above $m$th order `Lovelock' Riemann, one can project it in the plane orthogonal to $\ell _{a}$ and obtain the following scalar
\begin{equation}
\mathbb{R}=\mathtt{R}^{ab}_{cd}\mathbb{P}^{c}_{a}\mathbb{P}^{d}_{b}
\end{equation}
Explicit evaluation of the above scalar can be performed keeping in mind that the tensor $\mathtt{R}^{ab}_{cd}$ is antisymmetric under exchange of the indices $(a,b)$ and $(c,d)$ respectively, leading to
\begin{align}\label{Eq_Love_F_02}
\mathbb{R}&=\mathtt{R}^{ab}_{cd}\delta ^{c}_{a}\delta ^{d}_{b}-\frac{2}{\ell ^{2}}\mathtt{R}^{ab}_{cd}\ell^{c}\ell_{a}\delta ^{d}_{b}
\nonumber
\\
&=P^{ab}_{cd}R^{cd}_{ab}-\frac{(m-1)}{(d-1)(d-2)}(d^{2}-d)L
-\frac{2}{\ell ^{2}}\left\lbrace P^{ab}_{mn}R^{mn}_{cb}-\frac{(m-1)}{(d-1)(d-2)}\left(d\delta ^{a}_{c}-\delta ^{a}_{c}\right)L \right\rbrace \ell_{a}\ell^{c}
\nonumber
\\
&=mL-\frac{d(m-1)}{d-2}L-\frac{2}{\ell ^{2}}P^{ab}_{mn}R^{mn}_{cb}\ell_{a}\ell^{c}+2\frac{m-1}{d-2}L=-\frac{2}{\ell ^{2}}P^{ab}_{mn}R^{mn}_{cb}\ell _{a}\ell ^{c}+L
\end{align}
Thus equating \ref{Eq_Love_F_02} to the matter energy density through $-(2\kappa/\ell ^{2})T_{ab}\ell ^{a}\ell ^{b}$ we finally obtain,
\begin{align}
P^{a}_{bmn}R_{c}^{bmn}\ell _{a}\ell ^{c}-\frac{1}{2}L\ell ^{2}=\kappa T_{ab}\ell ^{a}\ell ^{b}
\end{align}
For unit normalized time-like vectors, $\ell _{a}=u_{a}$ and $\ell ^{2}=-1$, leading to $E_{ab}u^{a}u^{b}=\kappa T_{ab}u^{a}u^{b}$, where $E_{ab}\equiv P_{a}^{~pqr}R_{bpqr}-(1/2)Lg_{ab}$ is the analogue of Einstein tensor in Lovelock gravity. On the other hand, for the null vectors we arrive at $\mathcal{R}_{ab}\ell ^{a}\ell ^{b}=\kappa T_{ab}\ell ^{a}\ell ^{b}$, where the tensor $\mathcal{R}_{ab}\equiv P_{a}^{~pqr}R_{bpqr}$ is the analogue of Ricci tensor in Lovelock gravity. Hence even in the case of Lovelock gravity, if one assumes that $E_{ab}u^{a}u^{b}=\kappa T_{ab}u^{a}u^{b}$ holds for all timelike observers, the field equations for Lovelock gravity: $E_{ab}=\kappa T_{ab}$ follows. While in the null case, besides demanding the validity of $E_{ab}\ell^{a}\ell^{b}=\kappa T_{ab}\ell^{a}\ell^{b}$ for all null vectors, one have to use the Bianchi identity associated with Lovelock theories \cite{Padmanabhan:2011ex} as well as covariant conservation of matter energy-momentum tensor to arrive at the Lovelock field equations:  $E_{ab}+\Lambda g_{ab}=\kappa T_{ab}$, which inherits the cosmological constant as well. Note that the above result has been derived in the context of $m$th order \LL gravity, which can be generalized to the general \LL Lagrangian in a straightforward manner. This results into: $\{\sum _{m}E_{ab}^{(m)}\}+\{\sum _{m}\Lambda ^{(m)}\}g_{ab}=\kappa T_{ab}$. Interestingly, the cosmological constant besides being generated as an integration constant of the field equations also gets contribution from the $m=0$ term in the \LL gravity. Therefore one may choose this term in the Lagrangian appropriately to arrive at the present small value of the cosmological constant. Therefore we conclude that the derivation of field equations for gravity can always be achieved starting from the Poisson's equation and subsequently projecting a suitable curvature tensor, whether it is Einstein gravity or Lovelock. 
\section{Concluding Remarks}\label{Conclusion}

By equivalence principle gravity manifests itself by curving the spacetime, which the material particles follow. In particular one can invoke special relativity in locally freely falling frame and hence write down the laws of motion in curvilinear coordinates, thus describing motion in curved spacetime. The notion of locally freely falling observer brings in intrinsic observer dependence in the theory and introduces observers for whom a local spacetime region is causally inaccessible, known as a local Rindler observer. Remarkably the local vacuum state of a test quantum field (as fit for local inertial observers) will appear as thermal to the local Rindler observer \cite{Davies:1974th,Unruh:1976db} (however also see \cite{Lochan:2016cxt}). If any matter field (characterized by matter energy momentum tensor $T^{a}_{b}$) crosses the local Rindler horizon, it will appear to be thermalized by the Rindler observer (since the matter will take infinite time to reach the horizon) and the corresponding heat density is being given by $T^{b}_{a}\ell ^{a}\ell _{b}$ (for a perfect fluid the above quantity is given by $\rho+p$, which by Gibbs-Duhem relation is the matter heat density), where $\ell _{a}$ is the null normal to the horizon. Note that the above heat density for matter is invariant under the transformation $T^{a}_{b}\rightarrow T^{a}_{b}+(\textrm{constant})\delta ^{a}_{b}$.

At this stage, one can ask a natural question, ``what about heat density of gravity?''. Surprisingly, one can answer the same in the above setting. Considering a general null surface it turns out that one can interpret $R^{a}_{b}\ell ^{b}\ell _{a}$ as the heat density of the spacetime. This originates from the fact that one can have a one to one correspondence between $R_{ab}\ell ^{a}\ell ^{b}$ and the viscous dissipation term $2\eta \sigma _{ab}\sigma ^{ab}+\zeta \theta ^{2}$, where $\sigma_{ab}$ is the shear of the null congruence $\ell _{a}$ and $\theta$ is its expansion, with $\eta$ and $\zeta$ being shear and bulk viscous coefficients. Thus the term $R_{ab}\ell ^{a}\ell ^{b}$ is related to heating of the spacetime \cite{Chakraborty:2015hna,Padmanabhan:2010rp,Kolekar:2011gw,Bhattacharya:2017mrg,Bhattacharya:2015qkt}. Given this thermodynamic backdrop, it is clear that the Einstein's equations when written as \ref{Eq_02} not only yields the geometrical input of the gravitational theory but is also 
physically well-motivated since the equality of \ref{Eq_02} can be thought of as an equilibrium 
situation, where the heat produced by gravity is being compensated by that of matter. 

Thus the equation $2G_{ab}\ell ^{a}\ell ^{b}=T_{ab}\ell ^{a}\ell ^{b}$ not only arises more naturally from the relativistic generalization of Newton's law but the usefulness of writing Einstein's equations in this manner stems from the fact that one might interpret both sides of these equations \emph{independently} and the equations themselves follow due to a balancing act performed by spacetime itself \cite{Padmanabhan:2014jta,Padmanabhan:2015lla,Padmanabhan:2015pza,
Chakraborty:2015hna,Padmanabhan:2016eld,Wang:2015cna,Navia:2017ltw}. We would like to reiterate that the field equations derived in this context is purely geometrical and follows Newton's path. One first realizes that energy density associated with any material body is intrinsically observer dependent and surprisingly one can construct a tensor (again dependent on observer) which contains spatial derivatives of the metric alone. Keeping this as a curved spacetime generalization of Newton's law one uniquely arrives at Einstein's equations when one make use of all observers (or all the null surfaces). This shows that the most natural generalization of Newton's law to curved spacetime is $R_{ab}\ell ^{a}\ell ^{b}=8\pi T_{ab}\ell ^{a}\ell ^{b}$ (of course, leading to Einstein's equations, but at a secondary level) bolstering the claim that gravity is intrinsically a thermodynamic phenomenon.
\section*{Acknowledgements}

S.C. gratefully acknowledges the help of T. Padmanabhan for suggesting this project and also for helpful discussions throughout the same. He also thanks Naresh Dadhich, Kinjalk Lochan and Krishnamohan Parattu for various fruitful discussions. A part of this work was completed while the author was visiting Albert Einstein Institute in Golm, Germany and the author gratefully acknowledges the warm hospitality there. Research of S.C. is supported by the SERB-NPDF grant (PDF/2016/001589) from DST, Government of India.

\appendix
\labelformat{section}{Appendix #1} 
\section{An alternative Riemann tensor for Lovelock gravity}\label{App_01}

In Lovelock gravity it is possible to define two tensors having the symmetry properties of Riemann and satisfies Bianchi identity. The first one corresponds to defining a $(2m\times2m)$ tensor for $m$ th order Lovelock polynomial by multiplying $m$ such curvature tensors and then an alternating tensor of rank $(2m\times2m)$ as \cite{Kastor:2012se},
\begin{equation}
\mathcal{R}^{b_{1}b_{2}\ldots b_{2m}}_{a_{1}a_{2}\ldots a_{2m}}=\delta ^{b_{1}b_{2}\ldots b_{2m}}_{c_{1}c_{2}\ldots c_{2m}}\delta ^{d_{1}d_{2}\ldots d_{2m}}_{a_{1}a_{2}\ldots a_{2m}}R^{c_{1}c_{2}}_{d_{1}d_{2}}\cdots R^{c_{2m-1}c_{2m}}_{d_{2m-1}d_{2m}}
\end{equation}
Note that the above tensor is completely antisymmetric in both upper and lower indices. 

Let us now project all the indices on a lower dimensional spacelike hypersurface using the projection tensor $\mathbb{P}^{a}_{b}=\delta ^{a}_{b}-(1/\ell^{2})\ell^{a}\ell_{b}$, such that one obtains another $(2m\times 2m)$ tensor, but whose inner product with the normal $\ell_{a}$ identically vanishes. Hence one arrives at, 
\begin{equation}
m!~\mathbb{R}^{p_{1}p_{2}\ldots p_{2m}}_{q_{1}q_{2}\ldots q_{2m}}=\mathbb{P}^{p_{1}}_{b_{1}}\cdots \mathbb{P}^{p_{2m}}_{b_{2m}}\mathbb{P}^{a_{1}}_{q_{1}}\cdots \mathbb{P}^{a_{2m}}_{q_{2m}}\mathcal{R}^{b_{1}b_{2}\ldots b_{2m}}_{a_{1}a_{2}\ldots a_{2m}}
\end{equation}
One can immediately construct a scalar out of the projected $(2m\times 2m)$ tensor, leading to,
\begin{equation}
m!~\mathbb{R}=\mathbb{P}^{a_{1}}_{b_{1}}\cdots \mathbb{P}^{a_{2m}}_{b_{2m}}\mathcal{R}^{b_{1}b_{2}\ldots b_{2m}}_{a_{1}a_{2}\ldots a_{2m}}
\end{equation}
There would be two terms contributing to the above expression, one when all the projectors are related by Kronecker deltas and when one of the projector $\mathbb{P}^{a}_{b}$ is replaced by $\ell^{a}\ell_{b}$ while all the others are replaced by Kronecker deltas. In any other case, for example if two projectors are replaced by the normal, then it would identically vanish, thanks to the completely antisymmetric nature of $\mathcal{R}^{b_{1}b_{2}\ldots b_{2m}}_{a_{1}a_{2}\ldots a_{2m}}$. Thus finally we obtain,
\begin{align}
m!~\mathbb{R}&=\delta^{a_{1}}_{b_{1}}\cdots \delta^{a_{2m}}_{b_{2m}}\mathcal{R}^{b_{1}b_{2}\ldots b_{2m}}_{a_{1}a_{2}\ldots a_{2m}}
-\frac{2m}{\ell ^{2}}~\ell^{a_{1}}\ell_{b_{1}}\delta ^{a_{2}}_{b_{2}}\cdots \delta^{a_{2m}}_{b_{2m}}\mathcal{R}^{b_{1}b_{2}\ldots b_{2m}}_{a_{1}a_{2}\ldots a_{2m}}
\nonumber
\\
&=\delta ^{b_{1}b_{2}\ldots b_{2m}}_{c_{1}c_{2}\ldots c_{2m}}\delta ^{d_{1}d_{2}\ldots d_{2m}}_{b_{1}b_{2}\ldots b_{2m}} R^{c_{1}c_{2}}_{d_{1}d_{2}}\cdots R^{c_{2m-1}c_{2m}}_{d_{2m-1}d_{2m}}
-\frac{2m}{\ell ^{2}}~\ell^{a_{1}}\ell_{b_{1}}\delta ^{b_{1}b_{2}\ldots b_{2m}}_{c_{1}c_{2}\ldots c_{2m}}\delta ^{d_{1}d_{2}\ldots d_{2m}}_{a_{1}b_{2}\ldots b_{2m}} R^{c_{1}c_{2}}_{d_{1}d_{2}}\cdots R^{c_{2m-1}c_{2m}}_{d_{2m-1}d_{2m}}
\end{align}
One can now use the following identities,
\begin{equation}
\delta ^{b_{1}b_{2}\ldots b_{2m}}_{d_{1}d_{2}\ldots d_{2m}}\delta ^{c_{1}c_{2}\ldots c_{2m}}_{b_{1}b_{2}\ldots b_{2m}}=
\frac{m!}{2^{m}}\delta ^{c_{1}c_{2}\ldots c_{2m}}_{d_{1}d_{2}\ldots d_{2m}}
\end{equation}
as well as,
\begin{equation}
\delta ^{b_{1}b_{2}\ldots b_{2m}}_{c_{1}c_{2}\ldots c_{2m}}\delta ^{d_{1}d_{2}\ldots d_{2m}}_{a_{1}b_{2}\ldots b_{2m}} R^{c_{1}c_{2}}_{d_{1}d_{2}}\cdots R^{c_{2m-1}c_{2m}}_{d_{2m-1}d_{2m}}
=\frac{m!}{2^{m}}\delta ^{b_{1}}_{c_{1}}\delta ^{d_{1}d_{2}\ldots d_{2m}}_{a_{1}c_{2}\ldots c_{2m}} R^{c_{1}c_{2}}_{d_{1}d_{2}}\cdots R^{c_{2m-1}c_{2m}}_{d_{2m-1}d_{2m}}
\end{equation}
such that one arrives at,
\begin{align}\label{Eq_Love_F_01}
\mathbb{R}&=\frac{1}{2^{m}}\delta ^{d_{1}d_{2}\ldots d_{2m}}_{c_{1}c_{2}\ldots c_{2m}} R^{c_{1}c_{2}}_{d_{1}d_{2}}\cdots R^{c_{2m-1}c_{2m}}_{d_{2m-1}d_{2m}}
-\frac{2m}{\ell ^{2}}\frac{1}{2^{m}}\ell ^{a_{1}}\ell _{b_{1}}\delta ^{b_{1}}_{c_{1}}\delta ^{d_{1}d_{2}\ldots d_{2m}}_{a_{1}c_{2}\ldots c_{2m}} R^{c_{1}c_{2}}_{d_{1}d_{2}}\cdots R^{c_{2m-1}c_{2m}}_{d_{2m-1}d_{2m}}
\nonumber
\\
&=L-\frac{2}{\ell ^{2}}\ell _{c}\ell ^{d}R^{cp}_{qr}P^{qr}_{dp}=L-\frac{2}{\ell ^{2}}\mathcal{R}_{ab}\ell ^{a}\ell ^{b}
\end{align}
where, $\mathcal{R}_{ab}$ stands for the analogue of Ricci tensor associated with the Lovelock gravitational action. Hence this particular Lovelock Riemann tensor reproduces the Lovelock field equations for gravity as well if the procedure outlined above is being followed. The above exercise explicitly demonstrates the robustness of the idea presented here.
\bibliography{Gravity_1_full,Gravity_2_partial,My_References}

\bibliographystyle{./utphys1}
\end{document}